\documentclass[aps,prl,floatfix,twocolumn,superscriptaddress]{revtex4-1}
\usepackage{amsmath,amssymb}
\usepackage[utf8]{inputenc}
\usepackage{graphicx}
\usepackage{xcolor}
\usepackage{hyperref}
\hypersetup{colorlinks=true,citecolor={blue},linkcolor={blue},urlcolor={blue}}
\usepackage{float}

\newcommand{\mi}{\mathrm{i}}

\begin{document}

\title{Excitation of localized condensates in the flat band of exciton-polariton Lieb lattice}

\author{Meng Sun}
\affiliation{Center for Theoretical Physics of Complex Systems, Institute for Basic Science (IBS), Daejeon 34126, Korea}
\affiliation{Basic Science Program, Korea University of Science and Technology (UST), Daejeon 34113, Korea}

\author{I. G. Savenko}
\affiliation{Center for Theoretical Physics of Complex Systems, Institute for Basic Science (IBS), Daejeon 34126, Korea}
\affiliation{Basic Science Program, Korea University of Science and Technology (UST), Daejeon 34113, Korea}

\author{S. Flach}
\affiliation{Center for Theoretical Physics of Complex Systems, Institute for Basic Science (IBS), Daejeon 34126, Korea}
\affiliation{Basic Science Program, Korea University of Science and Technology (UST), Daejeon 34113, Korea}

\author{Y. G. Rubo}
\affiliation{Center for Theoretical Physics of Complex Systems, Institute for Basic Science (IBS), Daejeon 34126, Korea}
\affiliation{Instituto de Energ\'{\i}as Renovables, Universidad Nacional Aut\'onoma de M\'exico, Temixco, Morelos, 62580, Mexico}

\begin{abstract}
We propose a way to directly excite compact localized condensates in a nearly flat band of the exciton-polariton Lieb lattice by short Laguerre-Gaussian pulses, and investigate the dynamics of these condensates in the presence of repulsive polariton--polariton interaction and distributed losses in the lattice. 
The evolution of a low-density compact polariton condensate shows fast Rabi oscillations between its excitonic and photonic components, with slow beatings of the Rabi oscillation amplitude. 
Both oscillations and beatings are suppressed at higher condensate densities due to polariton-polariton repulsion and distributed losses in the lattice. A background incoherent pumping can be used to increase the lifetime and stability of compact localized states.
\end{abstract}

\date{\today}
\maketitle


\emph{Introduction.---}The full quench of a single-particle kinetic energy is the main feature of dispersionless or flat bands
\cite{Derzhko:2015aa,Leykam:2018aa,Leykam:2018ab}.
In many-body physics, it leads to a drastic manifestation of even weak interactions between particles. A prominent example of unusual fermionic correlations is the fractional quantum Hall effect showing itself in the flat Landau levels. 
Particles with bosonic statistics are also expected to dramatically change their properties in the flat band settings. Due to high degeneracy of the flat band energy level, one can construct compact localized states (CLSs), which extend over a few lattice sites only for a certain tight-binding model. The first such observation in a two-dimensional dice lattice is due to Sutherland \cite{Sutherland:1986aa}. 
If the concentration of bosonic particles is low, they can be distributed over several CLSs in such a way that their wave functions do not overlap, so that the total energy is minimized in the case of repulsive interaction between particles. As a result, depending on the number of occupied sites, the bosons can develop a supersolid phase, featuring periodic density modulation~\cite{huber10}. 

Can bosons with finite lifetime be loaded into a flat band and what are the expected effects in this case? We address this question using exciton-polaritons, that represent strongly coupled states of microcavity photons and semiconductor quantum well excitons~\cite{kavokin17}. Driven-dissipative condensates of exciton-polaritons have been reliably observed in semiconductor microcavities~\cite{kasprzak06,balili07}, and the potential of polariton condensates in artificial lattices for both applied and fundamental research has been intensively explored ever since. The $\pi$-condensates at the edges of bands in one-dimensional (1D) periodic potentials~\cite{lai07} and $d$-condensates in two-dimensional (2D) square lattices~\cite{kim11} have been demonstrated. There is now growing interest in exciton-polariton condensation in more complicated artificial periodic potentials, which target topologically protected~\cite{karzig15,nalitov15,bardyn15,stjean17,chunyanli18,solnyshkov18} and flat single-particle bands. Flat bands have been studied in honeycomb~\cite{jacqmin14}, kagome~\cite{masumoto12,gulevich16}, 1D Lieb~\cite{baboux16} and 2D Lieb~\cite{klembt17,whittaker18} lattices. The polariton condensates observed in flat bands are characterized by a rather short coherence length, and it is unclear whether this happens due to the potential disorder, or whether fragmentation is a generic feature of out-of-equilibrium condensation in flat bands.

In this Letter we consider a 2D Lieb lattice, with a geometry similar to Ref.~\cite{klembt17}. We investigate the combined effect of distributed dissipation and exciton-photon coupling on the miniband structure. First, by examining both the energy and lifetime of the particles, we identify possible candidate states for condensation in each miniband. We show that while there is no perfect flat band in this continuous, ``non-tight-binding'' system, the concept of long-lived strongly localized states, maintained by the destructive interference of propagating waves, is still valid to some extent. Secondly, we suggest a solution to the problem of cultivating compact localized condensates (CLCs) of exciton-polaritons and maintaining them for some operational time. The existing experimental ways to excite a polariton flat band utilize prolonged in space (cigar-shaped) incoherent pumps. The formation of the flat-band condensate requires fast relaxation time for particles to scatter down in energy. The flat band can be populated only under certain conditions in order to avoid condensation into different
bands. Moreover, multiple states get populated by the incoherent pump, decreasing the signal-to-noise ratio. We propose a resourceful way of exciting an exciton-polariton CLC by a resonant Laguerre-Gaussian pump targeted at particular placket of the Lieb lattice, and study the dynamics and evolution of the obtained condensate in the presence of losses and polariton-polariton repulsion. Finally, we show that in the presence of a background close-to-threshold incoherent pump, the CLC can be maintained for time intervals substantially exceeding the single-polariton lifetimes.

%
%
%
\begin{figure}[t!]
\includegraphics[width=0.99\linewidth]{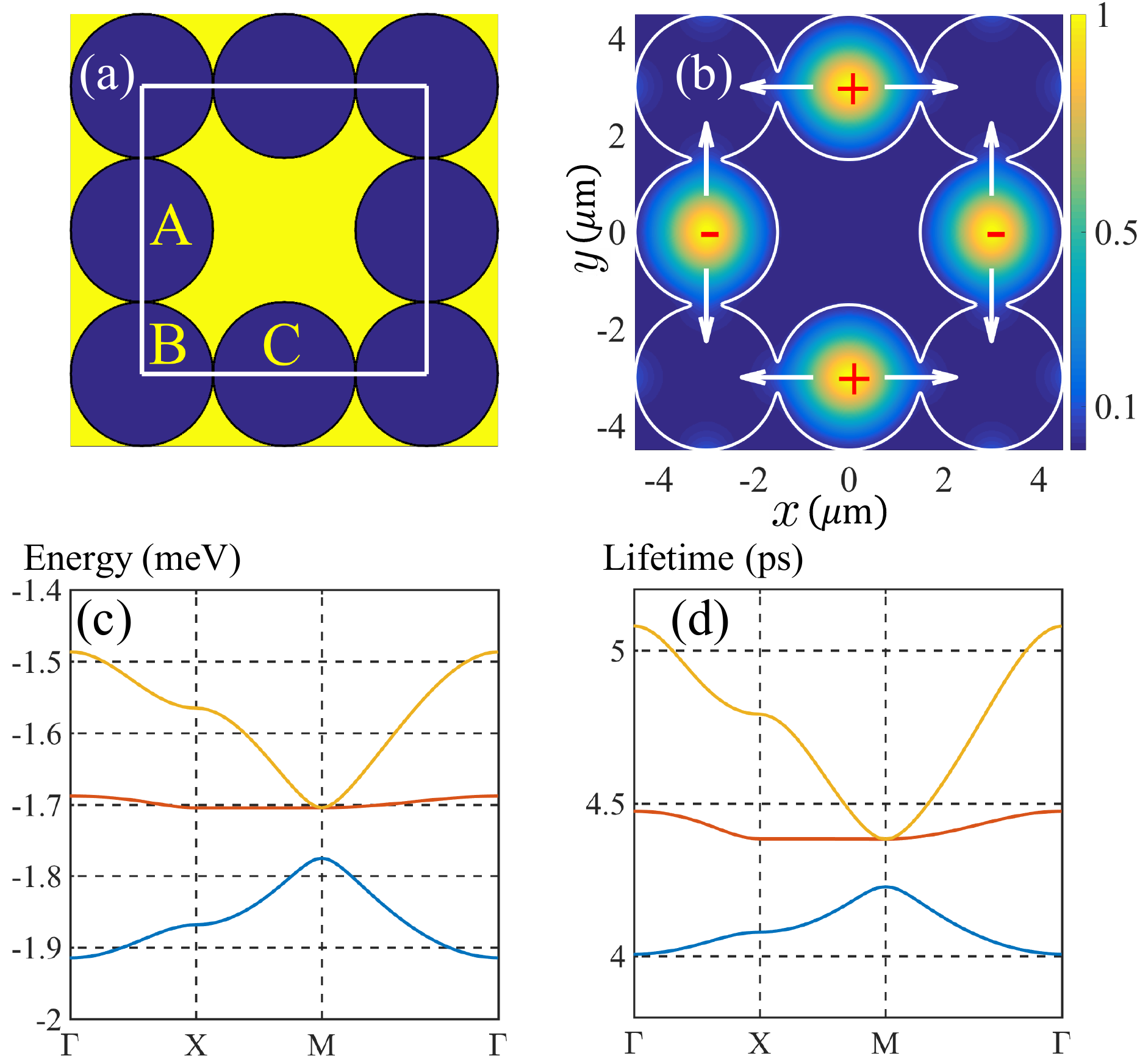}
\caption{System schematic (a-b) and single-particle spectrum (c-d). 
(a) The Lieb lattice placket which includes three pillars (quantum wells) per unit cell: A, B, and C.
(b) Probability density of the photonic component of the single-polariton (Bloch) state of the nearly flat (second) band at the $\Gamma$-point. 
Signs indicate the wave function phase. The weak population of the B sites is not visible.
A CLS possesses a similar structure and it will not propagate along the arrow directions due to destructive interference caused by the $\pi$-phase difference at sites A and C.
(c) The real part of the energy of the single-particle Bloch bands.
(d) The lifetimes of the Bloch states (the inverse imaginary part of the eigenvalues).
}
\label{fig:1}
\end{figure}
\emph{System schematic.---}The exciton-polariton condensate wave function can be written as $\Psi=(\varphi,\chi)^\mathrm{T}$, where $\varphi$ and $\chi$ are the photonic and excitonic components, respectively. 
Then, the mean-field Hamiltonian of the system reads (in what follows we set $\hbar=1$)
\begin{equation}\label{Ham}
  \hat{H} = \begin{pmatrix}
  -\frac{\nabla^2}{2m_c}+V(\mathbf{r}) & \frac{1}{2}\Omega \\
  \frac{1}{2}\Omega & \delta -\frac{\mi}{2\tau_x} -\frac{\nabla^2}{2m_x}
  +\alpha_x|\chi|^2
  \end{pmatrix},
\end{equation}
where $m_c$ and $m_x$ are the microcavity photon and exciton effective masses, respectively, $\Omega$ is the Rabi frequency, $\alpha_x$ is the exciton-exciton interaction strength, $\tau_x$ is the exciton lifetime, and $V(\mathbf{r})=V_r(\mathbf{r})-{\mi}V_i(\mathbf{r})$ is the complex-valued potential experienced by the photonic component separately from the excitonic component~\cite{sun2017multivalley}. The real part of the potential $V_r$ is defined by the quantum wells forming the Lieb lattice [see Fig.~\ref{fig:1}(a)], and the imaginary part $V_i$ describes the distributed losses in the system. 
We set $V_r=0$ and $V_i=0.1\,\mathrm{meV}$ inside the wells the lattice, while  $V_r=30\,\mathrm{meV}$ and $V_i=2.1\,\mathrm{meV}$ in the barriers. 
It should be noted, that the lifetime of photons is expected to be nonuniform. Indeed, the barriers are usually produced by partial etching of the distributed Bragg mirror, which introduces additional leakage of the photons from the barrier area. 
The diameter of each quantum well is chosen to be $3\,\mu\mathrm{m}$ and we also assume `touching' of the wells, so that the Lieb lattice constant is $a=6\,\mu\mathrm{m}$. The other parameters are $\hbar\Omega=9.5\,\mathrm{meV}$, $m_c=3.2\times10^{-5}\,m_e$, $m_x=10^5\,m_c$, $\tau_x=100$ ps, and we consider negative detuning of the microcavity mode with respect to the exciton mode, $\delta=-4.0\,\mathrm{meV}$. 

An elementary cell of the Lieb lattice is composed of three quantum wells, labeled as A, B, and C, as shown in Fig.~\ref{fig:1}(a). It is well known, that in the framework of a tight-binding model, the system spectrum possesses a flat band. The CLS in the tight-binding flat band is located on the A and C sites of the single placket. The phases on A and C are shifted by $\pi$, and the CLS is maintained due to destructive interference of waves propagating from the sites A and C to B~\cite{vicencio15}. 
The Bloch state of the nearly flat (second) band at the $\Gamma$-point, shown in Fig.\ref{fig:1}(b), has similar structure, except it is, of course, extended over the whole lattice. This state
also shows a $\pi$ phase shift between A and C sites, and in addition a very weak excitation of B sites.

%
Figure~\ref{fig:1}(c) shows the three lowest minibands in the system which represent the spectrum of noninteracting polaritons (i.e., for $\alpha_x=0$). 
Clearly, the continuous model described by the Hamiltonian~\eqref{Ham} does not lead to the appearance of a perfect flat band. 
The second miniband---flat within the tight-binding model with nearest-neighbor hopping---possesses a small, but finite dispersion. 

Another interesting and important feature of this system concerns the dispersion of losses in the bands shown in Fig.~\ref{fig:1}(d). 
For the lowest miniband, the state with the smallest losses occurs at the corner of the Brillouin zone (the M point) with the wave vector $k_x={\pm}k_y={\pm}\pi/a$. For the second (nearly flat) band, the minimal dissipation takes place at $k=0$ (the $\Gamma$ point). 
The wave function of this state corresponds to highly occupied A and C quantum wells and nearly empty B sites, as shown in Fig.~\ref{fig:1}(b). 

\begin{figure}[b!]
\includegraphics[width=0.99\linewidth]{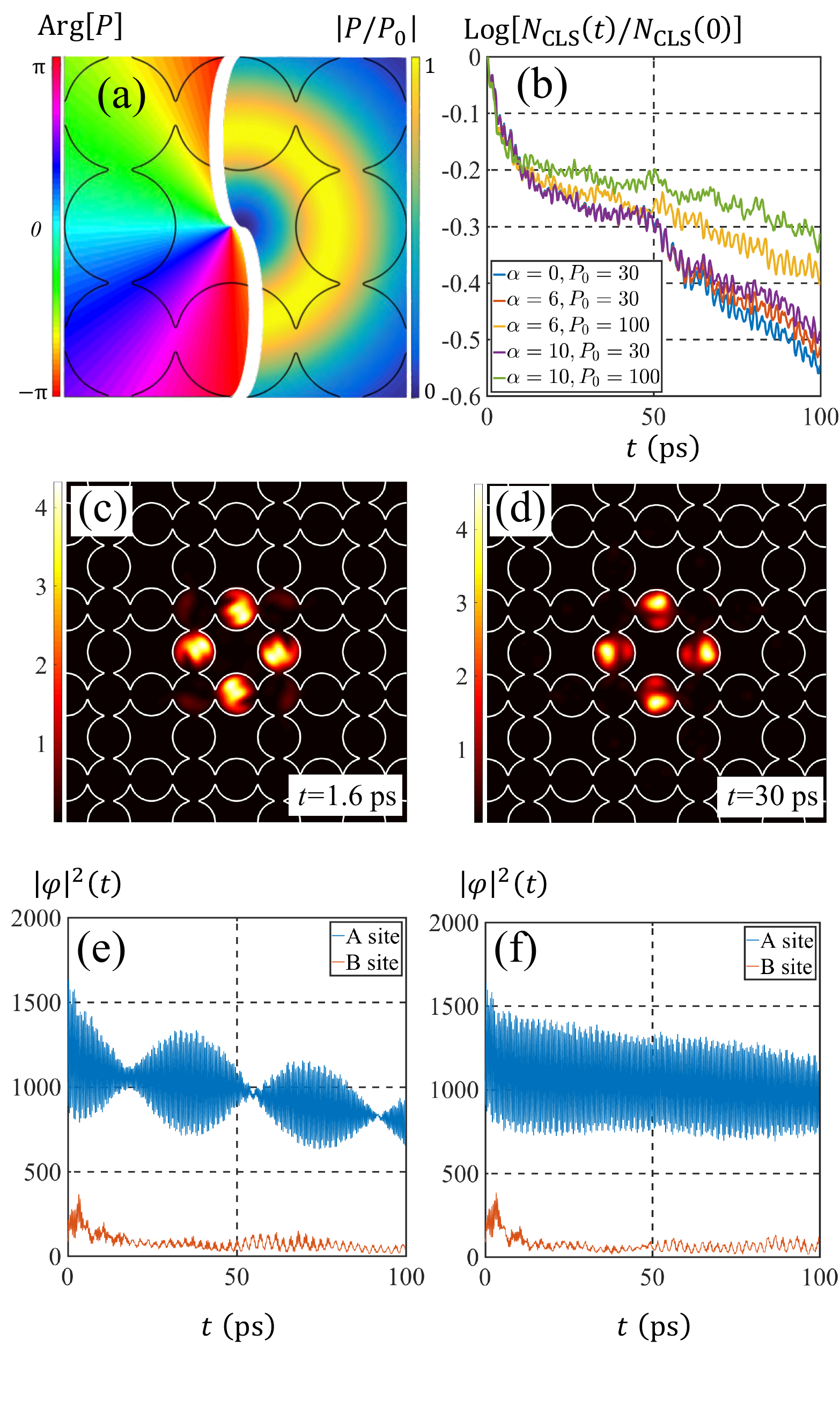}
\caption{(a) Phase and intensity of the Laguerre-Gaussian pulse with the radius $R=1.5\,\mu$m centered at the Lieb lattice placket. (b)  Decay of the CLC for different magnitudes of the interaction strengths $\alpha_x$ in the units of $\mu\mathrm{eV}\cdot\mu\mathrm{m}^2$ and the coherent strength $P_0$ in the units of $\mathrm{meV}\cdot\mu\mathrm{m}^{-1}$. 
(c-f) Dynamics of the CLC in the absence of dissipation.  
(c,d) The snapshots of the CLC particle density distribution, $\left(|\varphi(\mathbf{r},t)|^2+|\chi(\mathbf{r},t)|^2\right)d^2r$,
at two different times for $\alpha_x=10\,\mu\mathrm{eV}\cdot\mu\mathrm{m}^2$.
(e,f) the Rabi oscillations  of the photonic component from A and B sites for $\alpha_x=0$ (e) and $\alpha_x=10\,\mu\mathrm{eV}\cdot\mu\mathrm{m}^2$ (f) with $P_0=100\,\mathrm{meV}\cdot\mu\mathrm{m}^{-1}$.
}
\label{fig:2}
\end{figure}

\emph{The Laguerre-Gaussian resonant pump.---}We propose to excite the compact localized condensate (CLC) of the second band (which we refer to as the flat band in what follows) by exposing the Lieb lattice structure to a short, resonant, ring-shaped Laguerre-Gaussian pulse centered at one placket. The polariton wave function in this case evolves according to the equation
\begin{equation}\label{Evol}
  \mi\dot{\Psi} = \hat{H}\Psi+\begin{pmatrix}P(\mathbf{r},t)\\{0}\end{pmatrix},
\end{equation}
where the photonic pulse profile is given by~\cite{kim99}
\begin{equation}\label{Pulse}
  P(\mathbf{r},t)=P_0 \frac{(x\pm{\mi}y)^2}{R^2} 
  \exp\!\left[-\frac{r^2}{R^2}-\mi\omega_0t\right]\!\theta(t)\theta(t_p-t).
\end{equation}
Here $P_0$ is the pulse amplitude, $R$ is the  radius of the pulse ring, $\omega_0$ is the frequency of the pulse coinciding with the frequency of the flat band at the $\Gamma$ point, $\theta(t)$ is the Heaviside step function, and $t_p$ is the pulse duration. 
We are aiming at creating the CLC shown in Fig.~\ref{fig:1}(b). The transport of polaritons to the B sites should be blocked due to the $\pi$-phase difference of the wave functions on A and C sites. 
The phase and intensity plot presented in Fig.~\ref{fig:2}(a) shows that we can achieve this $\pi$ phase difference by centering the pump beam at the center of the unit cell [the center of the white square in Fig.~\ref{fig:1}(a)].


\emph{Dynamics of the CLC.---}To characterize the CLC dynamics, it is convenient to use the functions
\begin{equation}\label{NCLS}
N_\textrm{CLS}\left(t\right)=N_c(t)+N_x(t)=\int_{\textrm{A,C}}\left(|\varphi|^2+|\chi|^2\right) d^2r
\end{equation}
that measure the total number of particles residing at the cites A and C of the placket excited by the Laguerre-Gaussian pulse [shown in Fig.~\ref{fig:1}(a) and~\ref{fig:2}(a)]. We trace the evolution of the system just after the pulse is switched off at $t=0$. 
Figure~\ref{fig:2}(b) shows the decay rate of particles in the CLC for different intensities of the interaction strength and coherent pumpings. 

A counterintuitive result that one can see from Fig.~\ref{fig:2}(b) is the decrease of the particle loss from CLC with increasing the polariton-polariton interaction strength $\alpha_x$, or equivalently, with increasing the coherent pumping amplitude $P_0$, which puts more particles in the condensate and elevates the role of interaction. Apparently, in spite of the repulsive nature of exciton-exciton interaction, it has the focusing effect on the CLC in the Lieb lattice. 

Apart from a gradual decay of the excited CLC, we observe fast Rabi oscillations of particle number and more complex short and long time dynamics. To highlight these effects arising from the two-component (exciton and photon) nature of polaritons and their continuous, ``non-tight-binding'' propagation, we also present the peculiarities of the CLC dynamics in the absence of dissipation ($V_i(\mathbf{r})=0$ and $\tau_x^{-1}=0$).
Figures~\ref{fig:2}(c) and (d) show snapshots of the particle density $\left(|\varphi(\mathbf{r},t)|^2+|\chi(\mathbf{r},t)|^2\right)d^2r$
at two different times, $t=1.6\,\mathrm{ps}$ and $t=30\,\mathrm{ps}$, respectively. Due to the shape of the Laguerre-Gaussian pulse, the condensates excited in the A and C wells are smaller than the well size. These condensates bounce against the wells boundaries 
with a period $\sim 2\,\mathrm{ps}$, see Supplemental Materials~\cite{SM} (videos of this motion). 

Another interesting effect is slow modulation of the amplitude of the Rabi oscillations of the photonic component, which is usually measured experimentally~\cite{dominici14}. 
Figures~\ref{fig:2}(e,f) shows the time dependence of the total number of photons in an A site (the same as in a C site), as well as in a B site. For the interaction free case [Fig.~\ref{fig:2}(e)], one can see that the condensate dynamics at the A and C sites
is characterized by fast Rabi oscillations and a slow beating of their amplitude. 
The beating half-period $t_b$ is about $30\,\mathrm{ps}$, and it matches the width of the flat band ${\Delta}E_f\simeq0.02\mathrm{meV}\simeq\hbar/t_b$, so that the effect appears presumably due to the finite width of the band.
The beatings of the Rabi oscillations on the A(C) sites are suppressed and smeared out in the presence of polariton-polariton interaction [Fig.~\ref{fig:2}(f)]. It should be noted that the B sites occupation is very low.
%
%


\emph{Maintaining the CLC.---}The lifetime of polaritons in etched microcavities is typically short, making it hard to keep and operate the CLC for times longer than several ps. 
It follows from Fig.~\ref{fig:1}(d) that the lifetime of particles in the CLS (second band at $\Gamma$ point) is $\tau_\textrm{CLS}\approx4.5$ ps. One way to increase the operation time would be to use microcavities with higher quality factors. However, the losses can also be compensated by an incoherent background pumping, utilized to maintain the CLC.  
When the incoherent background pumping is present, the evolution of the system is described by the equations
\begin{subequations}\label{INC}
\begin{align}
 \mi\begin{pmatrix}\dot{\varphi}\\\dot{\chi}\end{pmatrix} &= 
 \hat{H}\begin{pmatrix}{\varphi}\\{\chi}\end{pmatrix}
 +\frac{{\mi}cn_r}{2}\begin{pmatrix}{0}\\{\chi}\end{pmatrix}
 +\begin{pmatrix}{P(\mathbf{r},t)}\\{0}\end{pmatrix}, \\[5pt]
 \dot{n}_r &= I - \tau_r^{-1} n_r-c|\chi|^2n_r,
\end{align}
\end{subequations}
where $n_r$ is the density of reservoir particles, $\tau_r=10\,\mathrm{ps}$ is their lifetime, $c=0.005$ ps$^{-1}\mu$m$^2$ is a phenomenological reservoir-system coupling rate, and $I$ is the intensity of the homogeneous incoherent pumping. 
To avoid excitation of polaritons in the first, the third and higher minibands,
we consider the intensity $I$ to be below the polariton condensation threshold.
In what follows, we use as a reference the threshold intensity $I_\mathrm{th}=(c\tau_r \tau_x)^{-1}$ for the excitonic component as a lower bound. 
\begin{figure}[t!]
\includegraphics[width=0.99\linewidth]{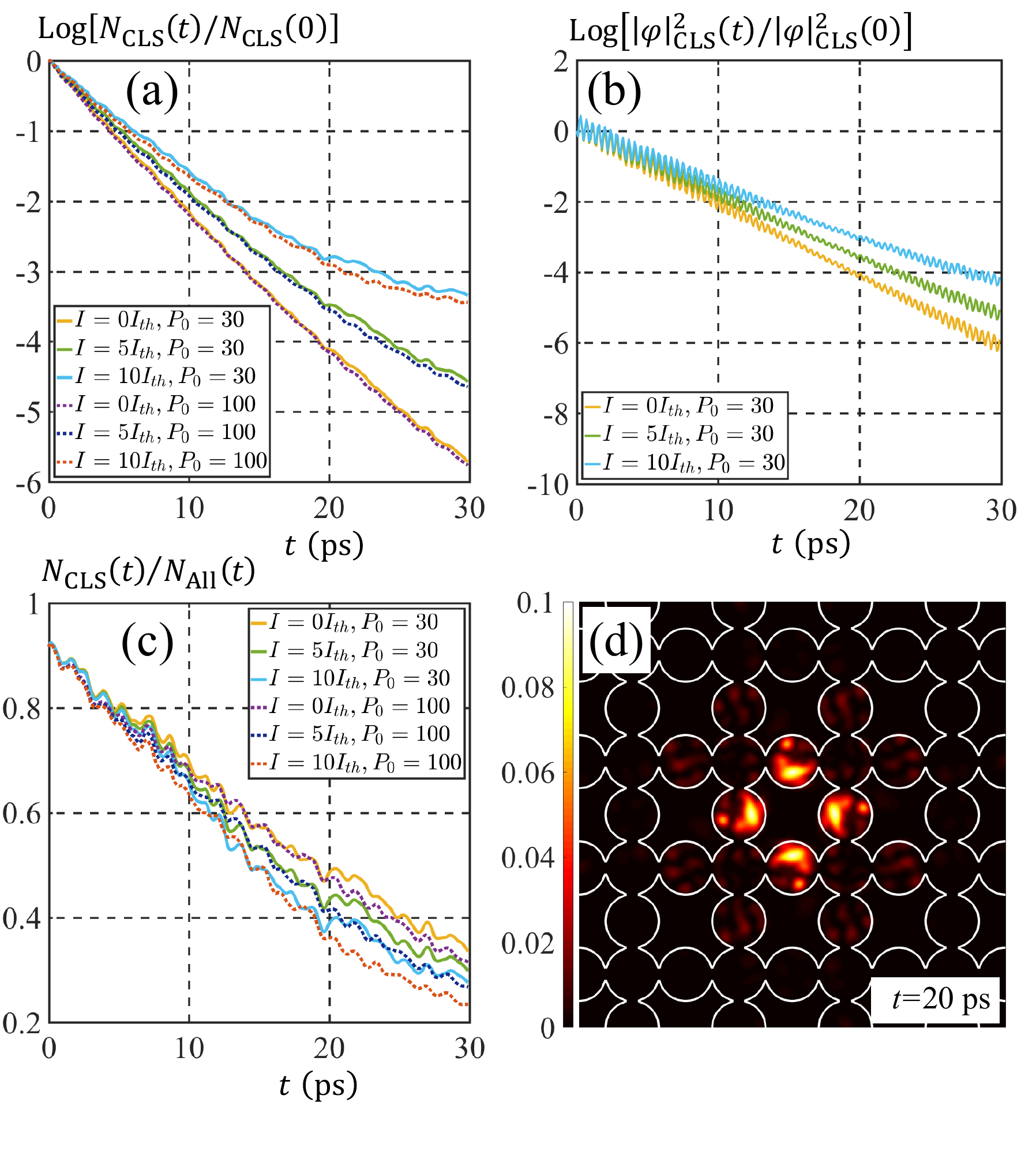}
\caption{(a) Decay of the CLC for different incoherent pumping intensities $I$ (and various coherent pumpings $P_0$). The Laguerre-Gaussian resonant pulse radius is $R=1.5~\mu$m. (b) Photonic decay of the CLC for different incoherent pumping intensities. (c) Evolution of the ratio of the CLC particles and the total number of particles for different incoherent pumping intensities. (d) Snapshot of the particle density in the CLC at 20~ps. }
\label{fig:3}
\end{figure}

Figure~\ref{fig:3}(a) shows the decay of particles residing in the CLC for different intensities of (both incoherent and coherent) pumping, together with the reference curve of the decay at $I=0$. The increase of $I$ does compensate the decay of particles from the CLC. 
The corresponding photonic decay also shows a similar behavior, as it is shown in Fig.~\ref{fig:3}(b). One can see from both panels that the Rabi oscillations persist in the presence of background incoherent pumping, indicating that the CLC maintains the coherence.

There are several shortcomings of the use of incoherent pumping. First, it leads to the excitation of 
particles in other (non-flat) bands and thus increases the occupation of the B sites. Secondly, although the background pumping allows maintaining the CLS for longer times, the price to pay is the generated noise. Figure~\ref{fig:3}(c) shows the ratio of particles in the CLC to the total number of particles in the system. The larger $I$, the worse is the single-to-noise ratio. At $I=10I_\mathrm{th}$ and after 20~ps, about $60\%$ of polaritons already left the CLC. However, even though the four CLC quantum wells contain only $40\%$ of polaritons, they still remain the most populated wells. That is following from Fig.~\ref{fig:3}(d) 
which shows a snapshot of the polariton condensate occupation density. Thus we conclude that the background pumping allows to keep the CLC for times which  one order of magnitude larger than the single-polariton lifetimes.

The coherent excitation of compact polariton condensates opens new possibilities to use the polariton Lieb lattice as a platform for network computations. In particular, it permits to construct graphs of compact localized condensates, similarly to recent proposals for classical~\cite{berloff17,ohadi17} and quantum~\cite{liew18} simulators. Both phase and polarization of localized condensates can be used to encode information.
The main benefits of the flat band states in the Lieb lattice consist of their compactness and suppressed in-plane spreading, 
as well as in better control of a multiple CLC arrangement, where all distances between CLCs are set by the underlying Lieb structure.     

\emph{In conclusion}, using an example of realistic two-dimensional exciton-polariton Lieb lattice with distributed losses, we have shown that the (nearly) flat band in this system possesses small but finite dispersion, both in the energy and the lifetime of the states. We have demonstrated the possibility to excite compact localized condensates in this nearly flat band using resonant Laguerre-Gaussian pulses. In spite of small dispersion of the band, the localization and coherence of compact localized condensates remain well defined. They exhibit an unusual dynamics, manifested by modulated fast Rabi oscillations. The coherent compact localized condensates can be maintained for times much longer than the polariton lifetime in the presence of an incoherent homogeneous background pumping.

We thank Alexey Andreanov, Carlo Danieli, Hugo Flayac, and Sukjin Yoon for useful discussions.
The authors acknowledge the support of the Institute for Basic Science in Korea (Project No.~IBS-R024-D1). YGR acknowledges support from CONACYT (Mexico) under the Grant No.\ 251808.

\bibliographystyle{apsrev4-1}
\bibliography{liebcite}

\end{document}